\documentclass[usenatbib,usegraphicx]{mn2e}
\usepackage{ctable}
\usepackage{amsmath}
\usepackage{amssymb}
\usepackage{float}

\title[Foregrounds for 21 cm Redshift Space Distortions]{The Impact of Foregrounds on Redshift Space Distortion Measurements
With the Highly-Redshifted 21~cm Line}

\author[J. C. Pober]{Jonathan C. Pober$^{1,2}$
\vspace{0.2cm} \\ 
$^{1}${Physics Department, University of Washington, Seattle, WA}\\
$^{2}${National Science Foundation Astronomy and Astrophysics
Postdoctoral Fellow}
}

\begin{document}
\maketitle

\begin{abstract}
The highly redshifted 21~cm line of neutral hydrogen has become recognized
as a unique probe of cosmology from relatively low redshifts ($z\sim1$)
up through the Epoch of Reionization ($z\sim8$) and even beyond.  To date,
most work has focused on recovering the spherically averaged power spectrum
of the 21~cm signal, since this approach maximizes the signal-to-noise in the
initial measurement.  However, like galaxy surveys,
the 21~cm signal is affected by redshift
space distortions, and is inherently anisotropic between the 
line-of-sight and transverse directions.  A measurement of this
anisotropy can yield unique cosmological information, potentially even
isolating the matter power spectrum from astrophysical effects.  
However, in interferometric measurements, 
foregrounds also have an anisotropic footprint
between the line-of-sight and transverse directions: the so-called
foreground ``wedge".  Although foreground subtraction techniques are
actively being developed, a ``foreground avoidance" approach of simply
ignoring contaminated modes has arguably proven most successful to date.
In this work, we analyze the effect of this foreground anisotropy
in recovering the redshift space distortion signature in 21~cm measurements
at both high and intermediate
redshifts.  We find the foreground wedge
corrupts nearly all of the redshift space signal for even the largest proposed
EoR experiments (HERA and the SKA), making cosmological information
unrecoverable without foreground subtraction.  The situation is somewhat
improved at lower redshifts, where the redshift-dependent mapping
from observed coordinates to cosmological coordinates significantly reduces
the size of the wedge.  Using only foreground avoidance, we
find that a large experiment like
CHIME can place non-trivial constraints on cosmological parameters.
\end{abstract}


\section{Introduction}

The highly-redshifted 21~cm line of neutral hydrogen has become
recognized as a unique probe of cosmology and astrophysics.  Due to its
small optical depth, 21~cm line observations
are sensitive to emission from neutral hydrogen
over nearly all of cosmic history.
Depending on the frequencies at which observations are conducted,
the 21~cm line has the potential to
offer insight into the nature of dark energy at late times 
(cosmic redshift $z~\sim~1~-~3$), the formation of the first galaxies 
during the Epoch of Reionization ($z \sim 6 - 13$), the birth of the first 
stars during ``cosmic dawn" ($z \sim 15 - 30$), and possibly 
even the physics of inflation and the early universe through observations of 
primordial fluctuations during the cosmic ``dark ages" ($z \sim 30 - 200$).
For reviews of the 21~cm cosmology technique and the associated science
drivers, see \cite{furlanetto_et_al_2006}, \cite{morales_and_wyithe_2011},
\cite{pritchard_and_loeb_2012} and \cite{zaroubi_2013}.
   
High-redshift 21~cm observations are not without their challenges.
The combination of an inherently faint cosmological signal
with extremely bright astrophysical foregrounds 
necessitates a dynamic range that has never been achieved with radio
telescopes.  In the search for fluctuations in the 21~cm signal
during the Epoch of Reionization
(as opposed to ``global" experiments targeting the mean signal evolution),
several experiments are currently conducting long observing campaigns,
such as
the LOw Frequency ARray (LOFAR; \citealt{yatawatta_et_al_2013,van_haarlem_et_al_2013})\footnote{http://www.lofar.org}, 
the Donald C. Backer Precision Array for Probing the Epoch of Reionization
(PAPER; \citealt{parsons_et_al_2010,parsons_et_al_2014})\footnote{http://eor.berkeley.edu},
and the Murchison Widefield Array (MWA; \citealt{lonsdale_et_al_2009,tingay_et_al_2013,bowman_et_al_2013})\footnote{http://www.mwatelescope.org}.
Given the limited sensitivity of these experiments, several larger
next-generation instruments are being designed, including
the low-frequency Square Kilometre Array (SKA-low; \citealt{mellema_et_al_2013})\footnote{http://www.skatelescope.org} 
and the Hydrogen Epoch of Reionization Array (HERA; \citealt{pober_et_al_2014})\footnote{http://reionization.org}.  Experiments to measure
the matter power spectrum and baryon acoustic oscillations (BAO)
at redshifts $\sim 1 - 3$ are also planned or under construction,
including 
the Canadian Hydrogen Intensity Mapping Experiment (CHIME; \citealt{shaw_et_al_2014})\footnote{http://chime.phas.ubc.ca},
the Tianlai cylinder array \citep{xu_et_al_2014}\footnote{http://tianlai.bao.ac.cn},
and the BAO Broadband and Broad-beam experiment (BAOBAB; \citealt{pober_et_al_2013a}).

Given the faintness of the cosmological signal, initial experiments
are generally focused on measuring the Fourier space power spectrum
of the 21~cm emission.  In real space, the cosmological
signal is isotropic, which allows a three-dimensional Fourier space
measurement to be averaged in spherical shells to produce a higher
signal-to-noise one-dimensional power spectrum measurement.  
Since many experiments lack the sensitivity to measure the 21~cm signal
without this Fourier space averaging, most of the literature has focused
on spherically averaged measurements.  However, as with galaxy surveys,
the 21~cm signal is not measured in real space but in redshift space.
Peculiar velocities in the line of sight direction complicate the simple
mapping between redshift and distance
\citep{jackson_1972}, causing redshift space distortions
\citep{sargent_and_turner_1977,kaiser_1987} and breaking the isotropy of the observed signal.
Whereas the value of isotropic power spectrum only depends on the magnitude
of the Fourier mode $k$, redshift space distortions break the
symmetry between the transverse $k_{\perp}$ modes in the
plane of the sky and the line of sight $k_{\parallel}$ modes.
Redshift space distortions are often parameterized by $\mu$, the cosine of 
the angle
between $k_{\perp}$ and $k_{\parallel}$ for a given mode;
a principal goal of redshift space distortion measurements
is to determine how the power spectrum changes as a function of $\mu$.

Several studies have investigated the effects of redshift space
distortions in measurements of the 21~cm signal and its power spectrum
\citep{bharadwaj_and_srikant_2004,barkana_and_loeb_2005,
mcquinn_et_al_2006,lidz_et_al_2007,mao_et_al_2008,mao_et_al_2012,majumdar_et_al_2013,jensen_et_al_2013}.
One of the most exciting results of these studies
is that the contributions of various components to the
21~cm power spectrum during the Epoch of Reionization 
--- the detailed shape of which is determined by a complicated
combination of ionization fluctuations, density fluctuations, and their
cross-correlations --- may be separable using the distinct redshift space
signatures of each term.  While this picture is certainly complicated
by non-linear effects, the prospect of recovering the primordial
density power spectrum at $z\sim10$ motivates continued efforts to
understand and eventually recover the signal.

21~cm redshift space distortions are less
studied during the $z \sim 1-3$ epoch, but since the 21~cm signal
is expected to closely trace the matter power spectrum during this period,
the results from studies of galaxy surveys are generally applicable
(e.g. \citealt{fisher_et_al_1994,heavens_and_taylor_1995,papai_and_szapudi_2008}
and \citealt{percival_and_white_2009}).  Measurements of redshift space
distortions at these moderate redshifts probe the history of
cosmic structure formation and have the potential to distinguish
dark energy models from modified gravity theories 
\citep{song_and_percival_2009}.

Measurements of redshift space distortions with the 21~cm line therefore
offer unique cosmological information.  Several studies have found
that such measurements are not beyond the realm of possibility
for next-generation and even current 21~cm experiments
\citep{mao_et_al_2008,jensen_et_al_2013}.  To date, however, the anisotropy
of foreground emission between the transverse and line of sight directions
has not been considered in conjunction with redshift space distortion
measurements.
While foreground emission does not actually ``live" in the cosmological
Fourier space where the power spectrum is measured, the same analysis
used to convert redshifted 21~cm line frequencies to cosmological distances
provides a methodology for determining which Fourier modes are
contaminated by foregrounds. 
This mapping is complicated by the ``mode-mixing" effects of the
interferometers used to make 21~cm measurements.  The inherently
chromatic nature of these instruments introduces spectral (and therefore
$k_{\parallel}$) structure which varies as a function of
$k_{\perp}$.  Studies over the last few years have found that
this mode-mixing causes otherwise smooth-spectrum foregrounds to
occupy an anisotropic wedge-like region of cylindrical 
($k_{\perp},k_{\parallel}$) Fourier space
\citep{datta_et_al_2010,vedantham_et_al_2012,morales_et_al_2012,parsons_et_al_2012b,trott_et_al_2012,thyagarajan_et_al_2013}.
Without subtraction of foreground emission,
the effect of the wedge is to limit the range of $\mu$ that can be
measured by 21~cm experiments, potentially spoiling their ability
to measure the redshift space signatures of interest.
The goal of this paper is to therefore determine the impact of the wedge on
potential redshift space distortion measurements with 21~cm experiments
at both high $z\sim6-10$ and moderate $z\sim1-3$ measurements.

The structure of this paper is as follows.
We first consider Epoch of Reionization redshifts ($z\sim6-10$)
in \S\ref{sec:hiz}.  Specifically, we describe the redshift
space distortion signature in \S\ref{sec:hiz-rsd}, the foregrounds
in \S\ref{sec:hiz-foregrounds}, and present sensitivity predictions for both
current and future 21~cm EoR experiments in \S\ref{sec:hiz-sense}.
\S\ref{sec:loz} follows a similar form, with
\S\ref{sec:loz-rsd} describing the redshift space distortion signal,
\S\ref{sec:loz-foregrounds} describing the foregrounds,
and \S\ref{sec:loz-sense} presenting sensitivity predictions, but for
lower redshift intensity mapping experiments.\footnote{While 21~cm experiments at all redshifts effectively produce a low-SNR, low resolution map,
the name ``intensity mapping" has become synonymous with 
experiments in the $z\sim 1-3$ range, and the two descriptions
are used interchangeably in this work.}
We conclude in \S\ref{sec:conclusions} with a focus on the important
differences between the high and moderate redshift regimes.
Unless otherwise stated,
all calculations assume a closed $\Lambda$CDM universe with $\Omega_m = 0.27,\
\Omega_\Lambda = 0.73$, and $h = 0.7$.
 
\section{Epoch of Reionization Measurements}
\label{sec:hiz}

\subsection{The Redshift Space Distortion Signal}
\label{sec:hiz-rsd}

During the epoch of reionization, the 21~cm power spectrum receives
contributions from both density and ionization fluctuations.  Generally,
the ionization fluctuations dominate the total power, but they are unaffected
by redshift space effects because only the density perturbations source
gravitational infall.
Therefore, the density fluctuation power spectrum will vary with $\mu$,
while the ionization power spectrum will remain isotropic.
\cite{barkana_and_loeb_2005} propose using this distinct angular dependence to
separate the two components (and their cross-correlation) to potentially
measure the density power spectrum alone.
\cite{mao_et_al_2012} present a ``quasi-linear" extension (like 
\cite{barkana_and_loeb_2005}, they use linear theory for treating the density
and velocity fluctuations, but include non-linear effects in their
calculations of ionization fluctuations) to this analysis.  In their
approximation, the power spectrum of 21~cm brightness fluctuations can be
written as:

\begin{equation}
\label{eq:quasilin}
P_{21}(k,\mu) = P_{\mu^0}(k) + P_{\mu^2}(k)\mu^2 + P_{\mu^4}(k)\mu^4,
\end{equation}
where the three moments are:
\begin{equation}
P_{\mu^0} = \widehat{\delta T_b}^2 P_{\delta_{\rho_{\rm{HI}}},\delta_{\rho_{\rm{HI}}}}(k),
\end{equation}
\begin{equation}
P_{\mu^2} = 2\widehat{\delta T_b}^2 P_{\delta_{\rho_{\rm{HI}}},\delta_{\rho_{\rm{H}}}}(k),
\end{equation}
\begin{equation}
P_{\mu^4} = \widehat{\delta T_b}^2 P_{\delta_{\rho_{\rm{H}}},\delta_{\rho_{\rm{H}}}}(k),
\end{equation}
where $\delta T_b$ is the mean brightness temperature of the 21~cm signal
relative to the CMB, and $\delta_{\rho_{\rm{H}}}$ and 
$\delta_{\rho_{\rm{HI}}}$ are the fractional overdensities of neutral hydrogen
and ionized hydrogen relative to the cosmic mean, respectively.

To produce a simulated signal for our calculations, 
we use the publicly available \texttt{21cmFAST}\footnote{http://homepage.sns.it/mesinger/DexM\_\_\_21cmFAST.html} version 1.01 code
\citep{mesinger_and_furlanetto_2007,mesinger_et_al_2011}.
\texttt{21cmFAST} is a semi-numerical code that provides three
dimensional simulations of the 21~cm signal over relatively large
volumes (400~Mpc in the simulations used here).
We use all the fiducial values of the 21~cm code 
(see \citealt{mesinger_et_al_2011} and \citealt{pober_et_al_2014}
for a description of the relevant parameters)
and assume
that $T_{\rm spin} \gg T_{\rm CMB}$ for the entirety of the simulation.
Rather than use the simulated 21~cm brightness temperature cubes,
we use the separate ionization and density fluctuation output
cubes to construct a $P(k,\mu)$ power spectrum
using the quasi-linear approximation of Equation \ref{eq:quasilin}.
As several authors have found, this quasi-linear formula only provides
a good approximation to the 21~cm power spectrum at relatively high
neutral fractions $(x_{\rm HI} \gtrsim 0.3)$.  We use a 
fiducial 21~cm power spectrum
with a neutral fraction of $x_{\rm{HI}}$ at $z \sim 0.5$.
We have confirmed that at this neutral fraction, 
the quasi-linear approximation produces a 21~cm power spectrum 
that is $\sim 25\%$ 
higher than the full non-linear calculation done by \texttt{21cmFAST}.
Since our predicted sensitivities in \S\ref{sec:hiz-sense} are quite poor,
using the quasi-linear approximation has the effect of being a conservative
error and has little effect on our conclusions.

\begin{figure}
\centering
\includegraphics[width=3.5in]{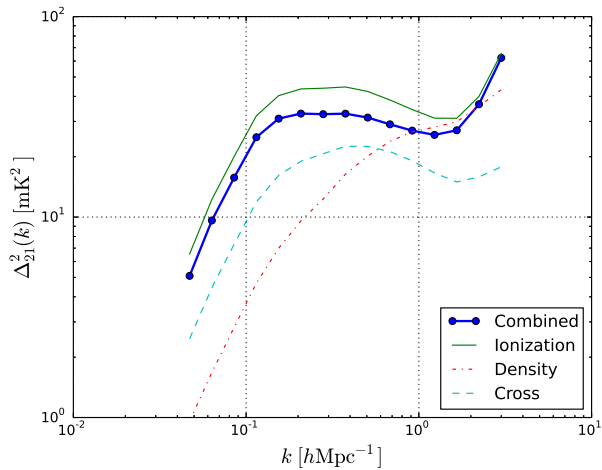}
\caption{The spherically averaged dimensionless power spectrum of our fiducial
model constructed using the quasi-linear approximation of Equation
\ref{eq:quasilin} (solid blue).  The individual components are shown as
thinner lines. 
The combined quasi-linear power spectrum is less than that of the ionization
power spectrum because the correlation between the ionization and density
fields, $P_{\delta_{\rho_{\rm{HI}}},\delta_{\rho_{\rm{H}}}}(k),$
 (dashed cyan line) is negative; its absolute value is plotted here.
This model was produced with simulations from
\texttt{21cmFAST}.} 
\label{fig:pspec1d}
\end{figure}

Figure \ref{fig:pspec1d} shows a spherically averaged version
of our fiducial quasi-linear $z = 9.5,\ x_{\rm HI} = 0.5$
power spectrum, where average values of $\left<\mu^2\right> = 1/3$ and 
$\left<\mu^4\right> = 1/5$ have been used in the averaging to account
for the effects of redshift space distortions.
Note that the overall 21~cm power spectrum is found to be fainter
than the ionization power spectrum, due to the negative cross-correlation
between the density and ionization fluctuations.  In general,
this correlation/anti-correlation can have a scale dependence and change sign
as a function of $|k|$, but in these simulations, the fluctuations
are anti-correlated at all scales.
Note that while 
one of the goals of redshift space distortion measurements will be to
measure their effects at a number of different redshifts, we use a single
redshift model for illustrative purpose here.  

\subsection{The Foreground Footprint}
\label{sec:hiz-foregrounds}

To model the effects of foregrounds in $(k_{\perp},k_{\parallel}$) space, we
use the approach of \cite{pober_et_al_2013a} and \cite{pober_et_al_2014},
where modes which fall inside the ``wedge" are deemed contaminated
and considered as if they were not measured.  We consider the wedge
as extending to the horizon limit \citep{parsons_et_al_2012b}, but do
not exclude any further
modes outside the horizon.  The mapping of the horizon
limit for a given baseline
(which is equivalent to the maximum delay between the two antennas
measured in light-travel time) to cosmological $k_{\parallel}$ modes is
given in \cite{parsons_et_al_2012b} and \cite{pober_et_al_2014}:
\begin{equation}
\label{eq:hor}
k_{\parallel,\rm{hor}} = \left(\frac{1}{\nu}\frac{Y}{X}\right) k_{\perp},
\end{equation}
where
$\nu$ is observing frequency, and $X$ and $Y$ are
cosmological scalars for converting observed bandwidths and solid
angles to $h{\rm Mpc}^{-1}$, respectively, defined in 
\cite{parsons_et_al_2012a} and \cite{furlanetto_et_al_2006}.\footnote{\cite{pober_et_al_2014} has a typographical error in which $X$ and $Y$ are reversed.}
The cosmological scalars $X$ and $Y$ are of particular importance here;
they depend on the angular diameter distance and Hubble constant
at the redshift of the measurement, and can change significantly
between the EoR experiments and the intensity mapping experiments
discussed in \S\ref{sec:loz}.  At the redshift of our fiducial
power spectrum $z=9.5$ the horizon limit gives a slope of
$k_{\parallel,\rm{hor}} = 3.73k_{\perp}$.  

\begin{figure}
\centering
\includegraphics[width=3.5in,clip=True,trim=7.5cm 0cm 0cm 0cm]{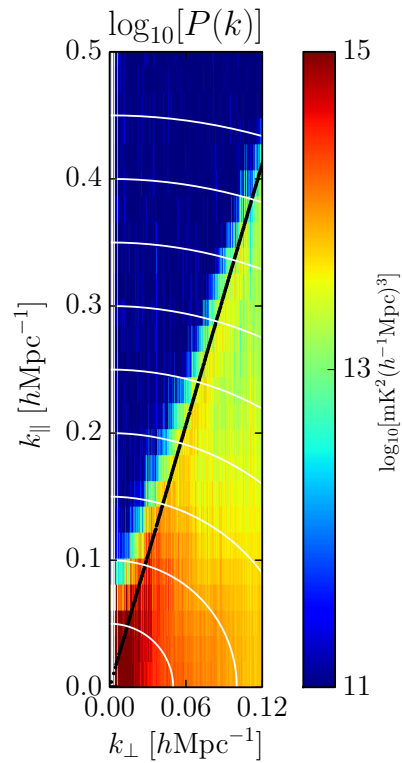}
\caption{The foreground wedge as observed in PAPER data from
\protect\cite{pober_et_al_2013b}.  This plot differs from their Figure 3, in that
the $k_{\parallel}$ and $k_{\perp}$ axes have been plotted in the same
scale.  The instrument measures much higher $k_{\parallel}$ modes than
shown, but the $k_{\perp}$ cutoff is real and set by the longest
baseline in the PAPER array.  The black line shows the analytic horizon
limit, and white lines show contours of constant $|k|$.}
\label{fig:datawedge}
\end{figure}

The steepness of this slope
is evident in Figure \ref{fig:datawedge}, which reproduces the wedge
obtained from PAPER observations in \cite{pober_et_al_2013a}\footnote{The observations of \cite{pober_et_al_2013a} are centered at 152.5~MHz,
where the horizon slope is 3.42, but the illustrative effect is unchanged.}.
Unlike \cite{pober_et_al_2013a}, here we plot 
the $k_{\parallel}$ and $k_{\perp}$
axes on the same scale to highlight another important
feature of EoR 21~cm experiments: a mismatch between $k_{\perp}$
and $k_{\parallel}$ scales.  The $k_{\perp}$ axis extends to 
$k_{\perp} = 0.12 h{\rm Mpc}^{-1}$, which corresponds to the longest
baseline in the PAPER array of $\sim 300$~m.  The $k_{\parallel}$ axis,
however, is
truncated; PAPER has a frequency resolution of 48.8~kHz, corresponding
to a maximum $k_{\parallel}$ of $10.4~h{\rm Mpc}^{-1}$. 
The white lines show contours of constant
$|k|$; even ignoring the wedge, a full range of $\mu$ is measured for
only very small values of $|k|$.
If modes within the wedge are considered contaminated, this loss of modes
serves to set a minimum $\mu$ below which modes cannot be measured.
For EoR observations at $z=9.5$, this value is $\mu_{\rm min} = 0.966$.
This large value of $\mu_{\rm min}$ is quite discouraging for EoR
experiments looking to measure the effects of redshift space distortions:
only with foreground removal working well into the wedge can a reasonable
range of $\mu$ be recovered. 

\subsection{Sensitivity Calculations}
\label{sec:hiz-sense}

Despite the discouragingly large value of $\mu_{\rm min}$ when modes
inside the wedge are excluded, there is still a small hope for
making a redshift space distortion measurement using only a foreground
avoidance approach. As Equation \ref{eq:quasilin} shows, 
the density power spectrum enters as $\mu^4$.  Therefore, even though
the measurable range of $\mu$ is extremely limited, it is at high
values of $\mu$, where the signal is changing most rapidly.  It is
therefore conceivable that for even measuring only modes outside the wedge,
an EoR experiment could pick out the component of the power spectrum
with a $\mu^4$ dependence with reasonable significance.  

In this section, we look at the potential for 21~cm experiments
to detect redshift space distortion effects in our fiducial power
spectrum.  To do so,
we use a version of the \texttt{21cmSense}
code\footnote{https://github.com/jpober/21cmSense} 
\citep{pober_et_al_2013a,pober_et_al_2014}
which has been modified to retain 2D ($k_{\perp},k_{\parallel})$
information (as opposed to the standard package, which performs
a spherical average to 1D).  
Under the foreground avoidance paradigm, 
all modes which fall inside the analytic horizon limit 
(Equation \ref{eq:hor}) are 
treated as foreground contaminted and excluded
from the sensitivity calculation.
This choice of exclusion region constitutes a middle ground 
between a more conservative choice where
an additive term is included to model ``supra-horizon'' emission
(c.f. Figure \ref{fig:datawedge}) and a more optimistic choice
where the primary field-of-view of the telescope is used instead
of the horizon (e.g. \citealt{beardsley_et_al_2012,pober_et_al_2014}).
While this may be a somewhat
pessimistic choice for next generation arrays with small fields
of view, it may also prove to be the case that 
foreground emission in primary beam sidelobes still overwhelms
the 21~cm signal in higher $k_{\parallel}$ modes.

To calculate the ability to recover redshift space distortion
information,
we fit a quartic polynomial in the
form of Equation \ref{eq:quasilin} to $P(k,\mu)$ for bins of constant
$|k|$. The constant term therefore recovers the isotropic 
ionization power spectrum at $|k|$,
the quadratic term the ionization-density cross power spectrum,
and the quartic term the density power spectrum.  Given the difficult
nature of the measurement, we consider two proposed next-generation
experiments: HERA \citep{pober_et_al_2014} and 
the core of Phase 1 of the SKA-Low
(following the design specifications of
\citealt{dewdney_et_al_2013})\footnote{Compared with the SKA
design used in \cite{pober_et_al_2014}, the model used here is spaced
out further by a factor of $\sim 3$ to better meet the specifications
of \cite{dewdney_et_al_2013}.}.  We summarize the properties of these
arrays in Table \ref{tab:eor_arrayinfo}.  We consider
observations spanning 1080 hours using the observing strategy
described in \cite{pober_et_al_2014}.
\begin{table*}
\centering
\begin{tabular}{c|p{.66in}p{.66in}p{.7in}p{3.3in}}
\hline
Instrument & Number of Elements & Element Size~$(\rm{m}^2)$ & Collecting Area~$(\rm{m}^2)$ & Configuration \\
\hline
HERA & 547 & 154 & 84,238 & Filled 200~m hexagon \\
SKA1-Low & 866 & 962 & 833,190 & Filled 270~m core with Gaussian distribution beyond \\
\hline
\end{tabular}
\caption{21~cm Epoch of Reionization Experiment Properties}
\label{tab:eor_arrayinfo}
\end{table*}

\begin{figure*}
\centering
\includegraphics[width=3.5in]{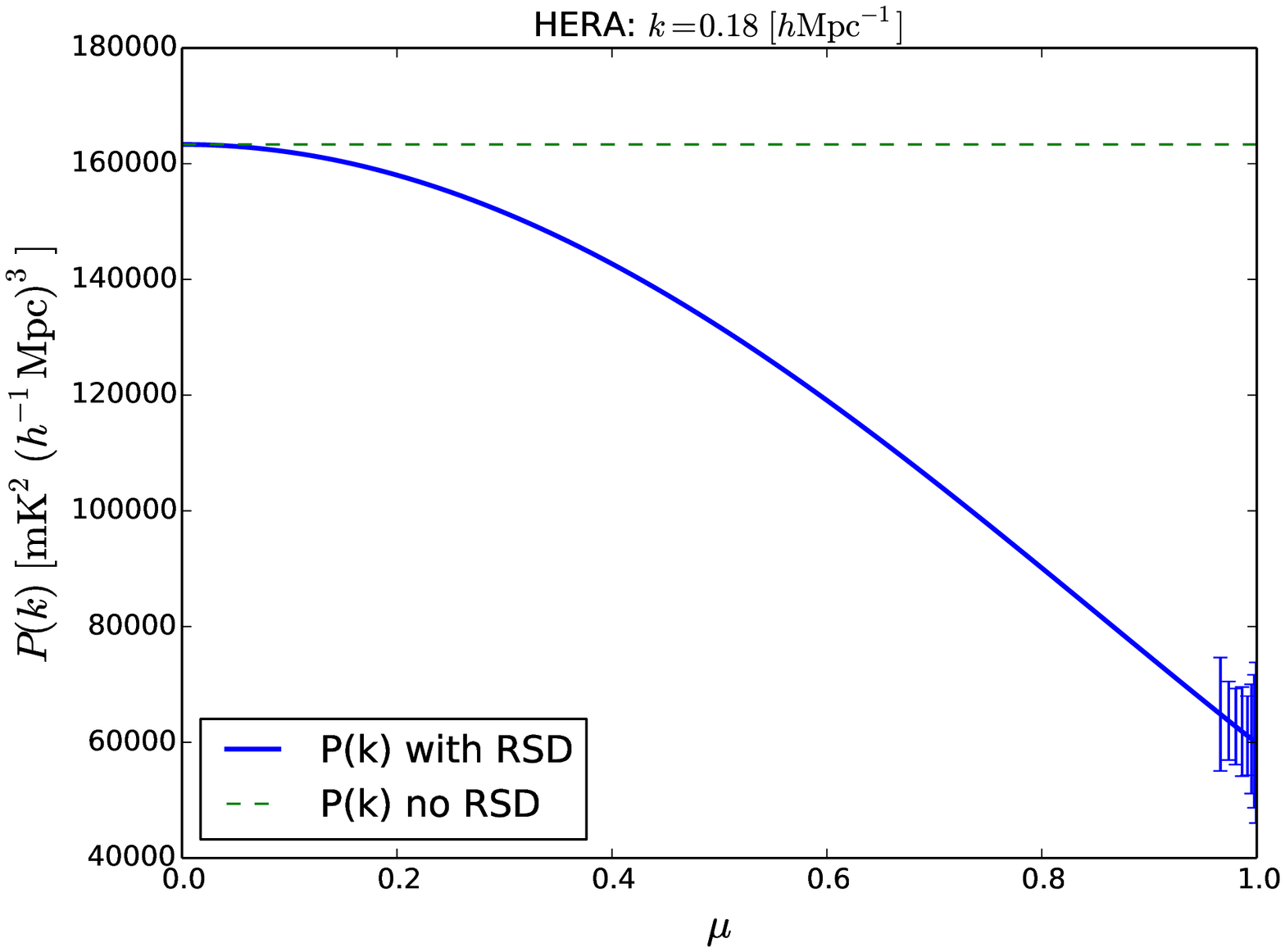}\includegraphics[width=3.5in]{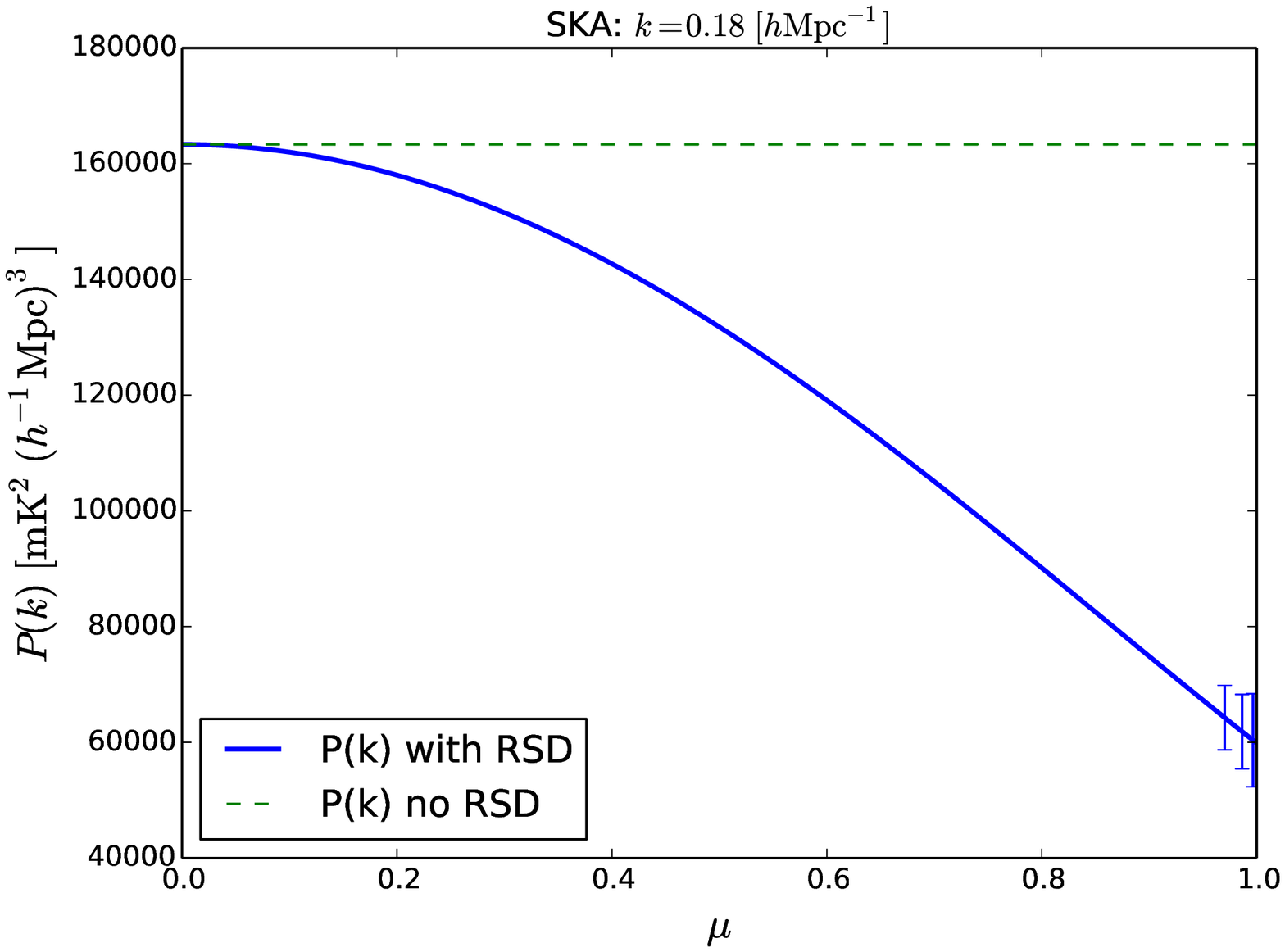}
\caption{Potential measurements of HERA (left) and the SKA (right) of 
the 21~cm power spectrum as a function of $\mu$ for $|k|=0.18~h{\rm Mpc}^{-1}$.
The blue line shows our fiducial 21~cm power spectrum including redshift
space distortion effects, while the green dashed line contains
only isotropic monopole term.  No binning of the measurements
has been performed; their spacing is set by the range of $k_{\perp}$
and $k_{\parallel}$ values probed by the instruments.
Only one value of $|k|$ is plotted, but the results are generic for all
$|k|$s: the foreground wedge limits measurements to $\mu > 0.97$, severely
hampering attempts to measure the redshift-space distortion signal
(i.e. to detect any $\mu$-dependence in the power spectrum).
}
\label{fig:pk_mu}
\end{figure*}

Figure \ref{fig:pk_mu} plots the power spectrum as a function of $\mu$ in
one annulus of $|k| = 0.18~h{\rm Mpc}^{-1}$.
\begin{table}
\centering
\begin{tabular}[hp]{c|p{0.35in}p{0.4in}p{0.4in}|l}
\hline
Instrument & Constant & Quadratic & Quartic & Spherically Avg.\\
\hline
HERA & 0.07 & 0.03 & 0.02 & 108.1 \\
SKA & 0.33 & 0.16 & 0.09 & 95.6
\end{tabular}
\caption{\rm First three columns: 
Detection significance (i.e. ``number of sigmas")
for each of the three $\mu$ moments
of the 21~cm power spectrum.  
Right hand column: Detection significance of the spherically
averaged power spectrum.
Despite very high SNR spherically averaged power
spectrum measurements, the $\mu$ dependence cannot be recovered
with any significance.}
\label{tab:eor_sense}
\end{table} 
The green curve shows the value
of the isotropic real-space power spectrum, while the blue curves shows
the effect of redshift space distortions as decreasing the power at high $\mu$,
the result of the negative sign of the density-ionization fluctuation cross
power spectrum.
At the $\mu$ values probed by EoR experiments, the decrease in
power is of order a factor of 2.  A more optimistic scenario might be to
consider a redshift where the cross power spectrum is positive and redshift
space distortions serve to boost the 21~cm signal.  However, as will
be shown below, the amplitude of the 21~cm power spectrum is not the limiting
factor, as both arrays deliver very high significance measurements
of the 1D spherically averaged power spectrum.
The left hand plot shows the measurements and associated
errors for HERA
in this $|k|$ annulus, while the right shows those for the SKA.

Table \ref{tab:eor_sense} shows the ``number of sigmas" at which
each experiment constrains the three components (constant, quadratic,
and quartic) of the redshift space power spectrum; while Figure
\ref{fig:pk_mu} shows the measurements only in one $|k|$ bin, these numbers
are for the cumulative measurement over all $|k|$s probed.
These significances are calculated by including the calculated errors
on each $k$ mode in the polynomial least-squares fit to the theoretical
power spectrum, yielding total errors on the fit coefficients
(which correspond to the three $\mu$ moments of the power spectrum).
Also listed are the total significance values for measurements
of the spherically averaged 1D power spectrum.
While the spherically averaged power spectrum is measured with very
high significance,
it is clear that even the next generation of 21~cm EoR experiments
will not make significant measurements of the redshift space distortion
effects without a foreground subtraction technique that allows
recovery of modes well inside the wedge.
It should also be noted that without a full range of $\mu$ measurements,
it will be difficult to separate the decrement of the power spectrum
amplitude at high $\mu$ from the isotropic amplitude.  
This error will potentially bias the interpretation of initial spherically
averaged power spectrum measurements, but 
attempting to quantify this effect is outside the scope of this present work.

\section{Intensity Mapping Experiments}
\label{sec:loz}

\subsection{The Redshift Space Distortion Signal}
\label{sec:loz-rsd}

The redshift space distortion signal is relatively simpler
at the $z\sim1-3$ measurements of intensity mapping experiments.
At these redshifts, all the neutral hydrogen resides in self-shielded halos,
which trace the matter power spectrum
\citep{madau_et_al_1997,barkana_and_loeb_2007}.  Since the entirety of the 21~cm
signal comes from only density fluctuations, there are no cross-correlation
terms between different components.  Furthermore, since 21~cm experiments
are probing mainly large scales that have not gone non-linear by these redshifts,
we use the Kaiser approximation to model redshift space distortions
\citep{kaiser_1987}:
\begin{equation}
\label{eq:rsd_loz}
P(k,\mu) = (1+ \beta \mu^2)^2 P(k),
\end{equation}
where $\beta \equiv f/b$ (where $f$ is the logarithmic growth rate of
structure $f \equiv d~\mathrm{ln}~D/d~\mathrm{ln}~a$ and $b$ is the bias of neutral hydrogen
containing halos). 
A principal goal of redshift space distortions measurements at these redshifts
is to constrain $\beta$, which has been measured to be $\approx \Omega_m^{0.6}$.
\citep{fisher_et_al_1994,heavens_and_taylor_1995,papai_and_szapudi_2008,percival_and_white_2009}.
A detailed history of $\beta$ as a function of redshift can be used to
trace the growth of structure over cosmic time and potentially
even distinguish between dark energy and modified gravity as the cause
of the observed late time acceleration in cosmic expansion
\citep{percival_and_white_2009,song_and_percival_2009}.
In 21~cm measurements, there is also a degeneracy in the power spectrum
amplitude between $b$ and $f_{\rm HI}$, the mass fraction of neutral
hydrogen with respect to the cosmic baryon content, 
which can be broken by measuring the $\mu$ dependence of the
power spectrum.

\begin{figure}
\centering
\includegraphics[width=3.25in]{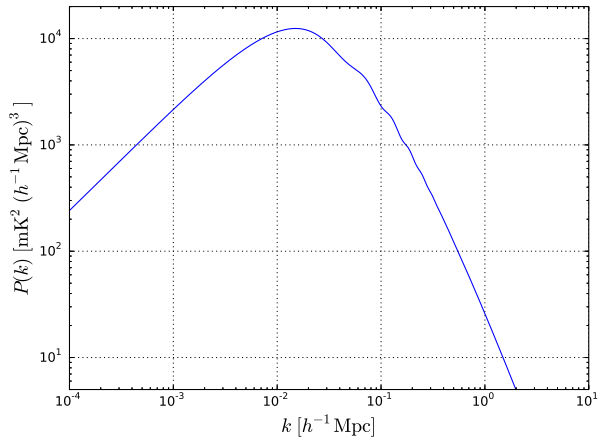}
\caption{The spherically averaged 21~cm power spectrum at $z=1.19$ including
redshift space distortion effects.}
\label{fig:pspec1d-loz}
\end{figure}

For our fiducial signal, we use a simulation of the matter power spectrum
at $z = 1.19$ from CAMB \citep{lewis_et_al_2000}\footnote{http://camb.info},
multiplied by a scalar converting the matter power spectrum
to 21~cm brightness temperature \citep{madau_et_al_1997,barkana_and_loeb_2007,ansari_et_al_2012,pober_et_al_2013a}:
\begin{equation}
    \label{eq:pred_sig_b}
    P(k) = \left[\tilde T_{21}(z)\right]^2 b^2 P_{\delta}(k),
\end{equation}
\begin{align}
    \label{eq:pred_sig_fhi}
    \tilde T_{21}(z) \simeq 0.084 {\rm mK} \frac{(1+z)^2 h}{\sqrt{\Omega_m(1+z)^3+\Omega_\Lambda}} \frac{\Omega_B}{0.044}\frac{f_{\rm{HI}}(z)}{0.01},
\end{align}
where $\tilde T_{21}(z)$ is the mean 21cm brightness temperature at
redshift $z$,
$P_{\delta}(k)$ is the matter power spectrum,
$\Omega_\Lambda$
is the cosmological constant, and $\Omega_m$ and $\Omega_B$ are the
matter and baryon density in units of the critical density,
respectively.  
Our fiducial value for $b$ is 1.5 \citep{chang_et_al_2010}.
We plot a 1D spherical average of our 
fiducial power spectrum,
including the redshift space effects of Equation \ref{eq:rsd_loz},
in Figure
\ref{fig:pspec1d-loz}.

\subsection{The Foreground Footprint}
\label{sec:loz-foregrounds}

Evaluating Equation \ref{eq:hor} for $z = 1.19$ (650~MHz in the 21~cm line,
near the center of band for a number of intensity mapping experiments)
gives a horizon slope of $k_{\parallel,\mathrm{hor}} = 0.77 k_{\perp}$,
a significantly shallower slope than at EoR redshifts.  This fact
mainly stems from the large evolution in the angular diameter distance
and Hubble parameter between $z \sim 8$ and $z \sim 1$. The corresponding
$\mu_{\rm min}$ for this horizon slope is 0.61; although a large range
of $\mu$ is still excluded by the wedge, there remain enough
high $\mu$ values that
measuring redshift space distortions at $z\sim1$ with a foreground
avoidance strategy becomes a feasible proposition. 

\subsection{Sensitivity Calculations}
\label{sec:loz-sense}
Following the procedure described in \S\ref{sec:hiz-sense}, we calculate
the sensitivities of three intensity mapping array concept designs
to redshift space distortion signatures in our fiducial $z = 1.19$ power
spectrum: a 144-element BAOBAB-like array \citep{pober_et_al_2013a},
a 4096-element CHIME-like array \citep{shaw_et_al_2014}, and an
SKA-Mid concept array meeting the design specifications of
\cite{dewdney_et_al_2013} (although for simplicity the 64 MeerKAT dishes
are assumed to be identical to the 190 SKA dishes).
\begin{table*}
\centering
\begin{tabular}{c|p{.66in}p{.66in}p{.7in}p{3.3in}}
\hline
Instrument & Number of Elements & Element Size~$(\rm{m}^2)$ & Collecting Area~$(\rm{m}^2)$ & Configuration \\
\hline
BAOBAB & 144 & 6.25 & 900 & Filled $12\times12$ square \\
CHIME & 4096 & 2.25 & 9,216 & Filled $64\times64$ square \\
SKA1-Mid & 254 & 176.6 & 44,863 & Random locations with power-law
baseline distribution \\
\hline
\end{tabular}
\caption{21~cm Intensity Mapping Experiment Properties}
\label{tab:loz_arrayinfo}
\end{table*}
\begin{figure*}
\centering
\includegraphics[width=2.25in]{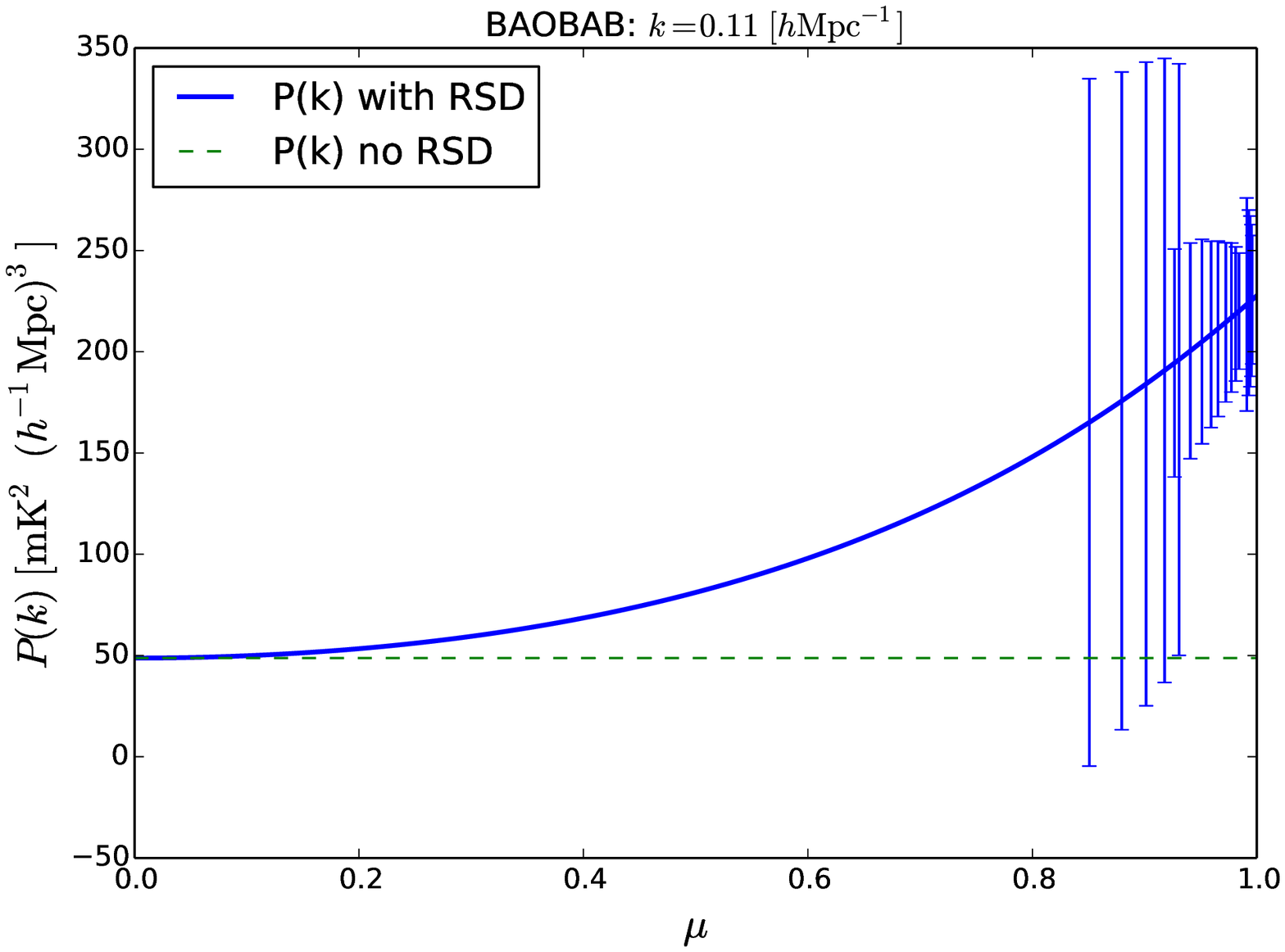}\includegraphics[width=2.25in]{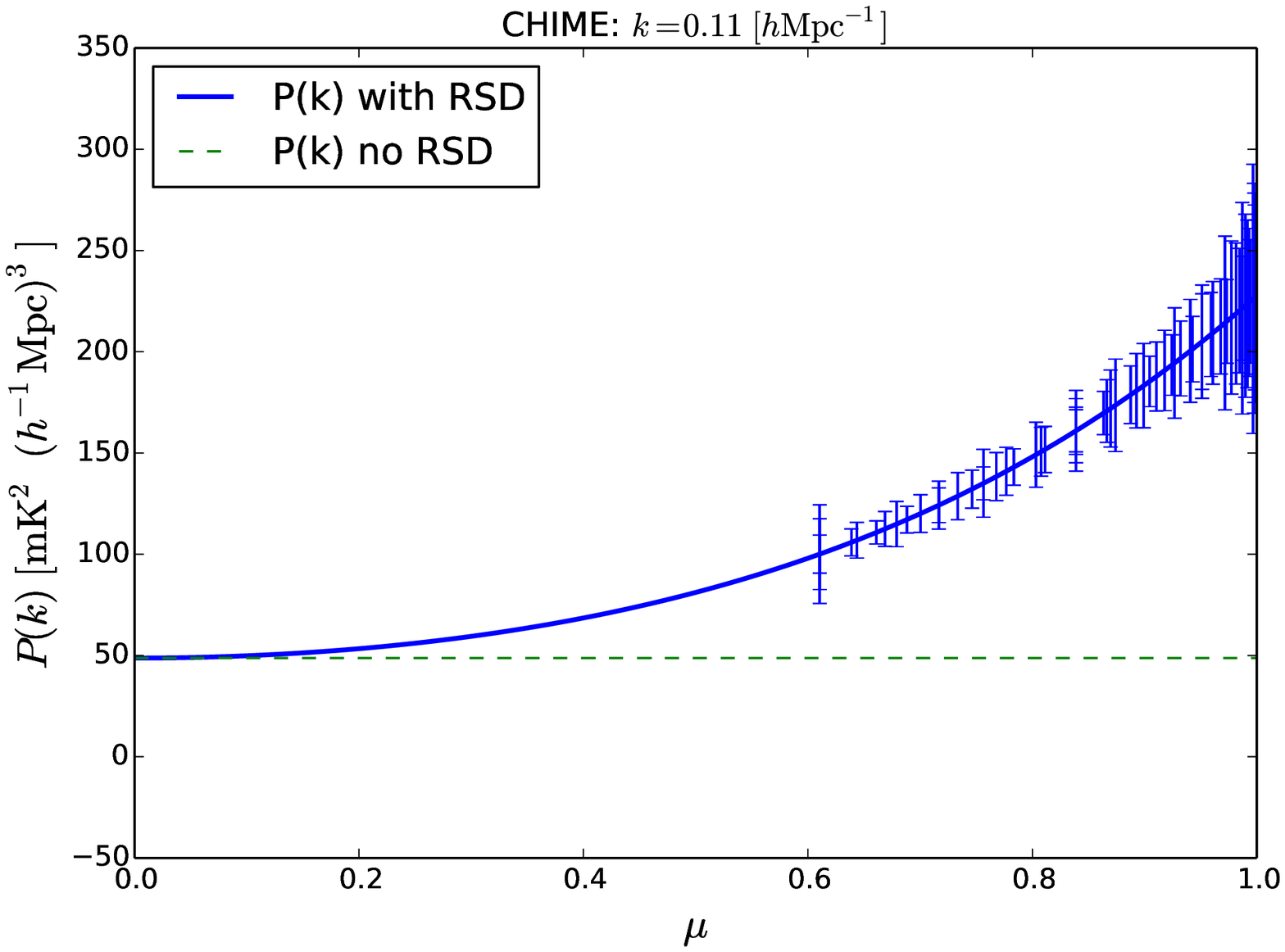}\includegraphics[width=2.25in]{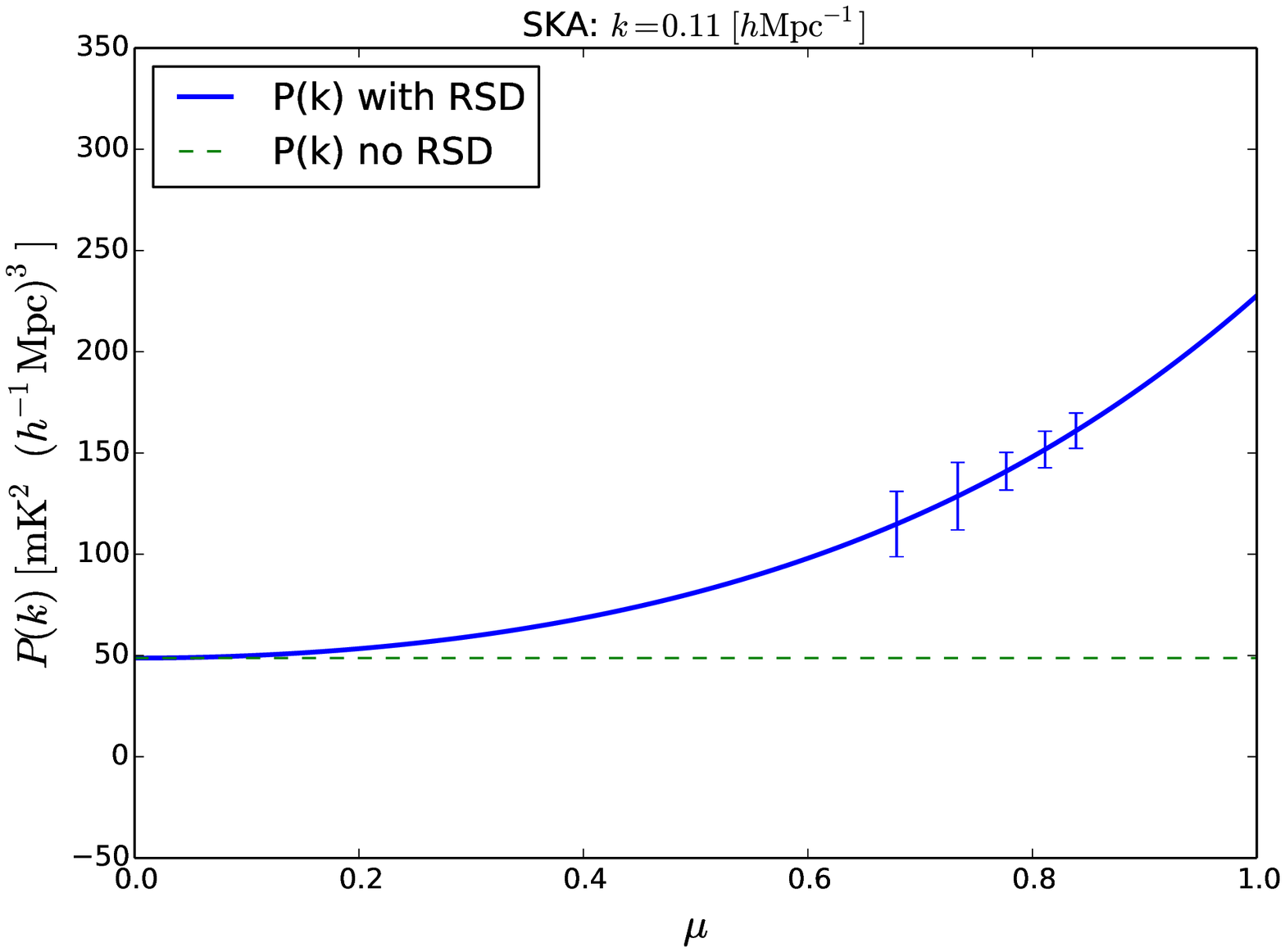}
\caption{Potential measurements of BAOBAB (left), CHIME (center), and 
the SKA (right) of 
the 21~cm power spectrum as a function of $\mu$ for $|k|=0.11~h{\rm Mpc}^{-1}$
(BAOBAB and CHIME) and $|k|=0.27~h{\rm Mpc}^{-1}$ (SKA).
The different $k$ mode and scale for the SKA results from the fact that
it cannot probe the shorter $k = 0.11~h{\rm Mpc}^{-1}$ mode
at 600 MHz.
The blue line shows our fiducial 21~cm power spectrum including redshift
space distortion effects, while the green dashed line contains
only isotropic monopole term.
No binning of the measurements
has been performed; their spacing is set by the range of $k_{\perp}$
and $k_{\parallel}$ values probed by the instruments.
Only one value of $|k|$ is plotted, but the results are generic for all
$|k|$s: low values of $\mu$ cannot be measured due to foregrounds, but
enough measurements are possible to see the $\mu$ dependence introduced
by redshift space distortions.  Only CHIME has enough SNR to fit
the functional form of the power spectrum and recover cosmological information.
} 
\label{fig:pk_mu_loz}
\end{figure*}
\begin{table*}
\centering
\begin{tabular}{c|lll|l|l}
\hline
Instrument & Constant & Quadratic & Quartic & Spherically Avg. & $\beta$ (frac. err.) \\
\hline
BAOBAB & 0.4 & 0.3 & 0.26 & 23.1 & 1.99 \\
CHIME & 5.7 & 4.3 & 3.4 & 205.8 & 0.15 \\ 
SKA & 1.8 & 1.4 & 1.1 & 65.66 & 0.47 \\
\end{tabular}
\caption{\rm Detection significance (i.e. ``number of sigmas")
for each of the three $\mu$ moments
of the 21~cm power spectrum, followed by
the total detection significance of the spherically averaged
21~cm power spectrum.  The right-hand column shows the achievable
fractional error on $\beta$.  Only CHIME produces measurements that
are of any cosmological significance.}
\label{tab:loz_sense}
\end{table*}

The configuration
details of these arrays are described in Table \ref{tab:loz_arrayinfo},
and their achievable constraints on the redshift space distortion signal
are presented in Table \ref{tab:loz_sense}.
For each array, we calculate the sensitivities for a 100~MHz band
centered on 650~MHz $(z = 1.19)$; most of these experiments have wider
bandwidths (e.g. $400-800$ MHz for CHIME), and so will be able to deliver 
measurements over a range of redshifts simultaneously.

We see that the relatively smaller BAOBAB-like array cannot make a significant
measurement of the redshift space distortion terms, while the CHIME-like
instrument can reach moderate significances.  The SKA design does 
poorly, despite its large collecting area; this is because
the minimum measurable $k_{\perp}$ for the 15~m dish of the SKA is
$0.07~h{\rm Mpc}^{-1}$.  This
scale is just beyond the peak of the power spectrum, 
and the vast majority of baselines are significantly longer.
Put another way, the SKA Mid design is not tuned for measuring
the large-scale structure of the neutral hydrogen power spectrum.
Example measurements at $k = 0.11\ h{\rm Mpc}^{-1}$ are shown for
the BAOBAB-like, CHIME-like, and SKA arrays 
in Figure \ref{fig:pk_mu_loz}.  As noted, the SKA has very few measurements
at this small value of $|k|$.
On the other hand, 
the BAOBAB-like array does not have measurements reaching down to 
$\mu_{\rm min}$ because it does not have enough long baselines
to measure $|k| = 0.11\ h{\rm Mpc}^{-1}$ modes with a significant
transverse component.  At smaller values of $|k|$ than plotted, 
the full accessible range of $\mu$ is recovered.

In Table \ref{tab:loz_sense}, we also present the achievable fractional
error on $\beta$, calculated
by taking ratios of different power spectrum moments
to cancel out uncertainties in other parameters and then propagating
theoretical uncertainties.  
Although more sophisticated analyses techniques now exist for redshift-space
distortion measurements (e.g. \citealt{percival_and_white_2009}), we present
constraints using simple calculations as illustrative of the instrument
sensitivities.  Presumably, more advanced analyses could be used to improve
these measurements.
Under this framework, a CHIME-like experiment can yield
15\% errors on $\beta$, comparable to the last generation of galaxy surveys
\citep{percival_et_al_2004,ross_et_al_2007}, while neither the smaller
BAOBAB-like instrument nor the SKA with its very long baselines
can make a significant measurement.
With significant constraints on $\beta$, an experiment can also break
the degeneracy between $\beta$ and $f_{\rm HI}$ in setting the amplitude of
the spherically averaged 21~cm power spectrum.
Finally, it is worth repeating that these measurement errors come from a
100~MHz band, meaning that the redshift dependence of $\beta$ over
the CHIME band should be measurable.   

To explore the broader potential of 21~cm experiments, we consider a
``no-foreground" case measurement with the CHIME-like array.  While
a completely foreground free measurement is impossible (the spectral
modes inherent in the foreground spectrum cannot be separated
from those modes in the 21~cm spectrum), this case serves to illustrate
the limitations of our foreground avoidance technique.
In this scenario, we obtain 2.5\% errors in a measurement of $\beta$.
While these errors could be further decreased by observing more fields
of view, it is clear that percent-level errors on $\beta$ will be
very difficult to achieve.

\section{Discussion and Conclusions}
\label{sec:conclusions}

In this paper, we have explored the effect of the redshift space
anisotropy of \emph{foregrounds} on measuring and recovering cosmological 
information from the redshift space anisotropy of the cosmological 21~cm
signal.  While a foreground avoidance approach has proven theoretically
\citep{parsons_et_al_2012b,pober_et_al_2014} promising and has yielded
the best upper limit of the EoR signal to date \citep{parsons_et_al_2014},
this technique severely limits the amount of information that can be
gleaned from redshift space distortions.  For 21~cm experiments at EoR
redshifts, the foreground wedge contaminates nearly all values of
$\mu = k_{\parallel}/|k|$; at 150~MHz, for example, all modes with
$\mu < 0.97$ will be corrupted by foregrounds without applying some foreground
removal/subtraction.  While measuring the $\mu^4$ moment of the 21~cm power
spectrum could potentially probe the matter power spectrum at high redshift,
the effect of the foreground-imposed $\mu$ cut-off is to prevent any
high-significance measurement even for planned large experiments
like HERA and the SKA.

The situation is somewhat more promising for lower redshift ``intensity
mapping" experiments.  While foregrounds still prevent recovery
of low $\mu$ modes, the window for making 21~cm measurements
is much larger.  This difference is due to the large evolution in the
Hubble parameter and angular diameter distance between redshifts
$z~\sim~8$ and $z~\sim~1$.  These cosmological parameters determine the
mapping of angular and frequency values in data to $k_{\perp},k_{\parallel}$ 
coordinates, with the effect of making the wedge much smaller at lower
redshifts.  In our calculations at $z=1.19$ over half of $\mu$ modes
still fall within the foreground wedge; however, since the
redshift space distortion signals are (to first approximation) 
quadratic and quartic in
$\mu$, larger values of $\mu$ are better for distinguishing the signal
from the $\mu$-independent monopole.  For a large 21~cm experiment
optimized for intensity mapping like CHIME, foreground avoidance will
still allow for the recovery of cosmological information from
the redshift space signal.  Smaller experiments like BAOBAB and less-optimized
experiments like the SKA, however, still cannot recover the signal.

While this work has primarily considered a
foreground avoidance technique, a ``foreground-free"
scenario was shown to increase intensity mapping constraints by as
much as $\sim 7$.  
While a no foreground scenario is clearly implausible,
further development of foreground
removal algorithms should allow for additional sensitivity in
redshift space distortion measurements with the 21~cm line.
Even if the end result of foreground subtraction is only to push the wedge
back from the horizon limit, this will have the effect of lowering
the minimum measurable $\mu$ mode, and so further open the window
on 21~cm redshift space distortion measurements.

\section*{Acknowledgements}
JCP is supported by an NSF Astronomy and Astrophysics 
Fellowship under award AST-1302774.  We thank 
James Aguirre, Daniel Jacobs, Adrian Liu and Matt McQuinn
for helpful conversations.
The data shown in Figure \ref{fig:datawedge} is the
result of the hard work of the entire PAPER collaboration.

\bibliographystyle{mn2e}
\bibliography{rsd}{}

\end{document}